# Protection of the patient data against intentional attacks using a hybrid robust watermarking code

Ahmad Nagm[1] and Mohammed Safy Elwan[2]

[1] Computer Engineering, Cairo Higher Institute for Engineering, Computer Science and Management, Cairo, Egypt
[2] Electrical Engineering, Egyptian Academy of Engineering and Advanced Technology, Cairo, Egypt

## ABSTRACT

The security of patient information is important during the transfer of medical data. A hybrid spatial domain watermarking algorithm that includes encryption, integrity protection, and steganography is proposed to strengthen the information originality based on the authentication. The proposed algorithm checks whether the patient's information has been deliberately changed or not. The created code is distributed at every pixel of the medical image and not only in the regions of non-interest pixels, while the image details are still preserved. To enhance the security of the watermarking code, SHA-1 is used to get the initial key for the Symmetric Encryption Algorithm. The target of this approach is to preserve the content of the image and the watermark simultaneously, this is achieved by synthesizing an encrypted watermark from one of the components of the original image and not by embedding a watermark in the image. To evaluate the proposed code the Least Significant Bit (LSB), Bit2SB, and Bit3SB were used. The evaluation of the proposed code showed that the LSB is of better quality but overall the Bit2SB is better in its ability against the active attacks up to a size of 2*2 pixels, and it preserves the high image quality.

## INTRODUCTION

The presence of electronic data of patients within the current worldwide health care information systems has important benefits for patients and care practitioners, including greater patient discretion and improved clinical management.

The flow of knowledge between hospitals, physicians, and others has increased by 40% between 2008 and 2009.

Hackers are also inspired by the growth and technical developments to intrude into the servers where these confidential data are stored. There are different kinds of attacks that could be used to manipulate these intelligent medical devices. If an attacker gets hold of a smart pacemaker, for instance, he will be able to give the patient a shock that could lead to his/her death (*Swarna Priya et al., 2020*).

The patient's information security and the patient's medical data privacy are of high importance nowadays. Medical information plays a significant role in the health system and its modification might result in misdiagnosis.

Medical patient information can be intentionally tampered with results being added or deleted.

Some image processing could also result in unintentional changes. For example, the loss in the patient's information while using the compression technique. This loss occurs in telemedicine applications to minimize the amount of information to be transmitted. Thus this process can induce unacceptable loss of knowledge depending on its degree which may result in a misdiagnosis.

Security requirements differ from application to application and the protected aspects that they underline. It will guarantee three characteristics: confidentiality, honesty, and availability (*Allaert & Dusserre, 1994*; *Armoni, 2000*; *Jennett et al., 1996*; *Katsikas, 2016*).

For the protection of the patient's information, digital watermarking techniques are used. Watermarking strategies can be categorized into spatial and frequency domain techniques in which the watermark is embedded.

Spatial watermarking has various techniques such as text mapping coding, Patchwork, Least Significant Bit, and Additive Watermarking, etc.

Compared to the transforming domain methods, the spatial domain methods are less complex but weak to various image attacks.

In this article, a spatial domain watermarking technique is proposed to improve the data payload and at the same time, both authentication and data hiding are included. For the identity authentication purpose, the created code is distributed into the region of interest and regions of non-interest parts of the cover medical image. To enhance the security of the confidential patient information, SHA-1 is used to get the initial Key for the Symmetric Encryption before embedding it in the medical image.

The main contribution of the present research is:

- The proposed approach preserves the patient data using a hyped watermarking strategy, which includes encryption, integrity protection, and steganography.
- The watermark is created from the image and each image has its watermark depending on one of the three components of the image and on the personal information of the patient.
- Its advantage over the other approaches is that the created code is distributed at every pixel of the cover medical image and not only in the regions of non-interest pixels (RONI).
- To enhance watermark security, the hash function is used to encrypt the watermark.
- Although there is a watermark in every pixel in the image, it represents 1/8 of the pixel, but it did not affect the quality of the image.
- It solves the problem that exists in the spatial domain algorithms that cannot simultaneously offer protection against intentional attacks and robustness.

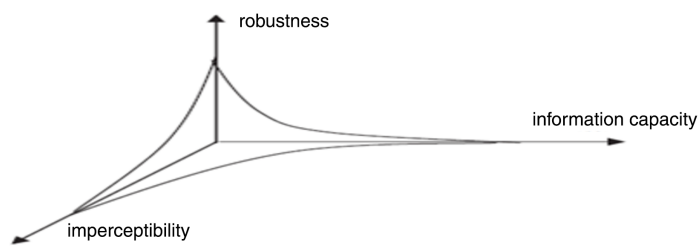

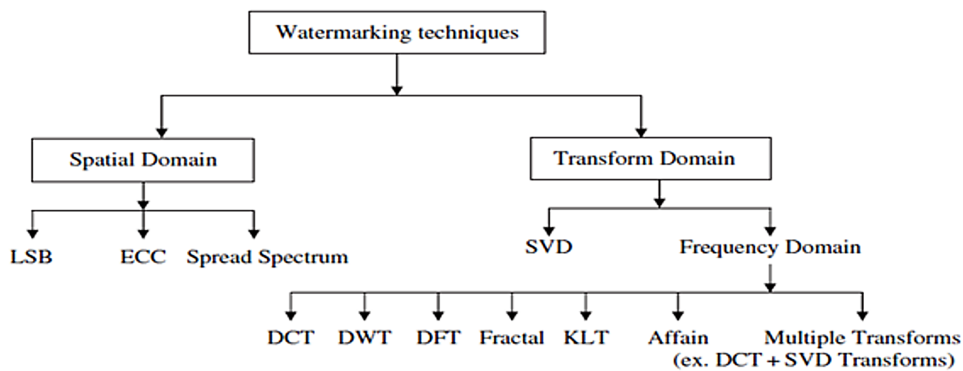

## LITERATURE REVIEW

To protect the medical images during transmission, watermarking is used. It is a process of embedding significant information over a patient's medical image to provide authentication, information hiding, tamper-proof data, etc. (*Pan et al., 2009*). The main factors of the watermarking process are reliability, confidentiality, and authenticity. The process of watermarking is classified into invisible and visible watermarking. For watermarking attacks (*Priya, Santhi & Swaminathan, 2012*), invisible watermarking is a robust technique.

In medical images, quality is critical because the diagnosis and treatment are the primary goals. For this reason, watermarking has several limitations in medical image authentication (*Rao & Kumari, 2011*). To create an efficient watermark the following criteria shall be met: (i) imperceptibility (ii) payload (iii) robustness against manipulations Fig. 1 shows the tradeoff between the watermark characteristics.

In *Mousavi, Naghsh & Abu-Bakar (2014)*, the quality requirements for patients' pathology medical data are extremely strict, and no changes are allowed. The research in the field of medical image digital watermarking (*Nyeem, Boles & Boyd, 2013*) is essential because any change in the transmitted patient information is forbidden and will affect the physician's decision.

Digital image watermarking can be mainly done in both transform and spatial domains as illustrated in Fig. 2.

In the spatial domain, the watermark is inserted within the original medical image (*Zhang & Wang, 2009*). Many techniques are used to insert a watermark such as LSB (Least Significant Bit) substitution technique, Pixel alteration, and bit shifting, etc.

The technique of the spatial domain watermarking is very easy and simple with less complexity. It is an easy task to embed watermarks into the spatial domain components of the cover images. It has been one of the basic schemes used since the beginning of digital watermarking in 1993 (*Pan, Huang & Jain, 2004*).

Usually, spatial watermarking systems choose several pixels from the cover image and adjust the luminance values of the pixels chosen depending on the bits of the watermark to be embedded (*Pan, Huang & Jain, 2004*; *Van Schyndel, Tirkel & Osborne, 1994*; *Tirkel et al., 1993*).

A well-known classic spatial-based watermarking method is the last-significant-bits (LSB) modification scheme (*Van Schyndel, Tirkel & Osborne, 1994*). The simplicity of implementation and the low complexity are the advantages of spatial-based watermarking schemes. On the other hand, they are not robust against common methods of attack (*Wolfgang & Delp, 1998*; *Voyatzis & Pitas, 1998*; *Langelaar, Setyawan & Lagendijk, 2000*; *Hernández, Perez-Gonzalez & Rodriguez, 1998*). To improve the spatial domain weakness, many methods are proposed.

In *Jennett et al. (1996)*, the performance of spatial-based watermarking schemes can be improved using the Bose–Chaudhuri–Hochquenghem (BCH) block codes.

In *Huang, Pan & Wang (2004)*, a scheme with greater robustness was proposed by the authors. Like the general scheme described, first of all, their scheme also selects the number of pixels considered from the cover image. Then, the mean value of the neighboring pixels is determined for each pixel calculated. This mean value is used to change the selected pixel. The robustness of their system is higher, which means that this system is more acceptable for practical use. Besides, the device also provides a parameter for regulating the balance between imperceptibility and robustness. Therefore, users of this scheme may prefer to have better imperceptibility or better robustness.

In *Memon & Gilani (2011)*, the content authentication of the patient's images (CT) is the main purpose. The image is separated into two regions one of them is the region of noninterest (RONI) and the other in the region of interest (ROI). The watermark is inserted in the RONI, so the quality of ROI has perseveted. This code is very simple but it can be easily attacked.

In *Zain & Clarke (2011)*, the security of the ultrasound (US) images is increased based on authentication and integrity. First, a rectangular shape is used to separate the RONI and the ROI. Then SHA256 hash function is used to calculate the hash value of the whole image. To make the code more secure, a secret key is used to create a hash value as well as a secret key for the inserted watermark. Finally, the hash value is inserted into the LSBs of RONI.

In *Wakatani (2002)*, data hiding is the main purpose. The digital watermarking is proposed to prevent the distortion of ROI in the original data by inserting the watermarking image in the RONI. The results show the high performance of the code to detect the embedded watermark, simultaneously the embedded watermark can be generated only apart from the ROI.

In *Viswanathan & Venkata Krishna (2011)*, watermarking and cryptography are used in one system in a way that inserts an encrypted version of the patient's original information. After watermarking, the corrupted details of the medical information were recovered using a reversible property.

In *Memon et al. (2011)*, a robust watermark is proposed based on the combination of the identification code of the physician, patient's information and, the LSBs of the ROI, which after encryption were embedded into the RONI of the patient's original image.

In *Nagm et al. (2019)* and *Nagm, Torky & Youssef (2019)*, a hybrid scheme is proposed. The code is distributed over the entire image. The code used the LSB, Bit4SB, and MSB. The main focus is to discuss the efficacy of the images and not focus on the ability to counter intentional attacks to alter the patient's information.

In *Alattar (2004)*, for color images, a reversible watermarking algorithm with very high data-hiding ability has been created. The algorithm enables the process of watermarking, which restores the exact original image, to be reversed. The algorithm obscures multiple bits in the differential expansion of neighboring pixel vectors. These findings show that at the highest signal-to-noise ratio, the spatial, quad-based algorithm allows for hiding the largest payload.

In *Shih & Wu (2005)*, based on the genetic algorithms the watermark is embedded around the ROI of the original image. The lossless compression is used to compress the ROI part and the lossy compression is used to compress the RONI part.

In *Thodi & Rodríguez (2007)*, solve the problem of the undesirable distortion at low embedding capacities of the Tian's difference-expansion technique. To embed the location map, the histogram shifting technique is used. A new reversible watermarking algorithm is illustrated based on the combination between the different expansion and the histogram shifting.

In *Zhao et al. (2011)*, the histogram theory is the core of this paper. Based on the difference between the pixels the histogram is constructed. A multilevel histogram modification mechanism is used in the embedding of the data. One or two level histogram modification is employed to improve the hiding capacity. The embedding level is used to extract the data and in the stage of the image recovery.

In *Wang, Lee & Chang (2013)*, based on the histogram-shifting a reversible data hiding approach is proposed. By changing the peak point pixel value to another pixel value in the same section, the hidden data can be inserted in the cover image. To guarantee the correct extraction of the secret data, the proposed method uses a location map.

In *Shih & Zhong (2016)*, the medical image is separated into ROI and RONI. The medical images with multiple ROIs preselected by experts are taken as the input, to improve the embedding ability without distorting the significant diagnostic information and keep ROIs lossless. The RONI is embedded with the watermark. Because of the image transformation being applied, the ROIs are confined to the rectangular shape.

In *Bamal & Kasana (2018)*, for the embedding of the watermark in the cover image the slantlet transformation along with the RS vector is used. For watermark security, MD5 and AES are used. The watermark is embedded in all the channels of the original image. The results show an increase in the capacity of the image and at the same time maintains the visual quality of watermarked images.

In *Savakar & Ghuli (2019)*, a combination of blind and non-blind watermarking is used. The watermark is a binary image and embedded in the cover image using DWT.

In *Al-Afandy et al. (2018)*, two-hybrid techniques are used. The first one is based on Discrete Stationary Wavelet Transform (DSWT) and the other one is based on the Singular Value Decomposition (SVD) in Discrete Wavelet transform. The results show that the YIQ and the RGB color coordinate system are robust in watermarking techniques against the attacks. On the other hand, the HSV is not robust in the embedding of the digital watermark.

In *Ahvanooey et al. (2020)*, to make the trace unnoticeable to the readers an invisible watermark is embedded in a cover text. Also, an instance-based learning algorithm is used to mark all the words of the original text. The result shows that the proposed algorithm has a higher performance compared with the existing codes. On the other hand, the algorithm uses the traditional way of embedding watermark.

To summarize the findings from the existing works, spatial watermarking has the advantage of being simple implementation and low complexity and on the other hand, it is not efficient against the attacks. All the existing algorithms rely on one strategy that is the separation of the original image into the region of interest and region of non-interest. The efficiency of the algorithms depends on their ability to recover the watermark from the image. In some algorithms, the rectangular shape of the image is a must.

This work is an extension to *Nagm & Safy (2020)* but the steganography process is carried out in the spatial domain. Through this work, we tried to prove the effectiveness of the proposed watermark in the face of the intentional alteration of patient data not only in the frequency domain but also in the spatial domain.

## PROPOSED METHODOLOGY

From all the above-mentioned purposes, a hybrid spatial domain watermarking scheme is proposed that meets the requirements of the medical imaging authentication.

The big difference between the proposed code and the above-mentioned researches is the distribution of the watermark. The watermark is distributed at every pixel in the image includes the region of interest and region of non-interest. Although the code satisfies the payload requirements because the changes in each pixel represent 1/8 from the pixel and according to the location of the bit.

Another difference is the target of the algorithm is to preserve the watermark and the patient image simultaneously, unlike other algorithms, which are its most important goal is to extract the embedded watermark. The whole structure of the proposed algorithm is shown in *Fig. 3*.

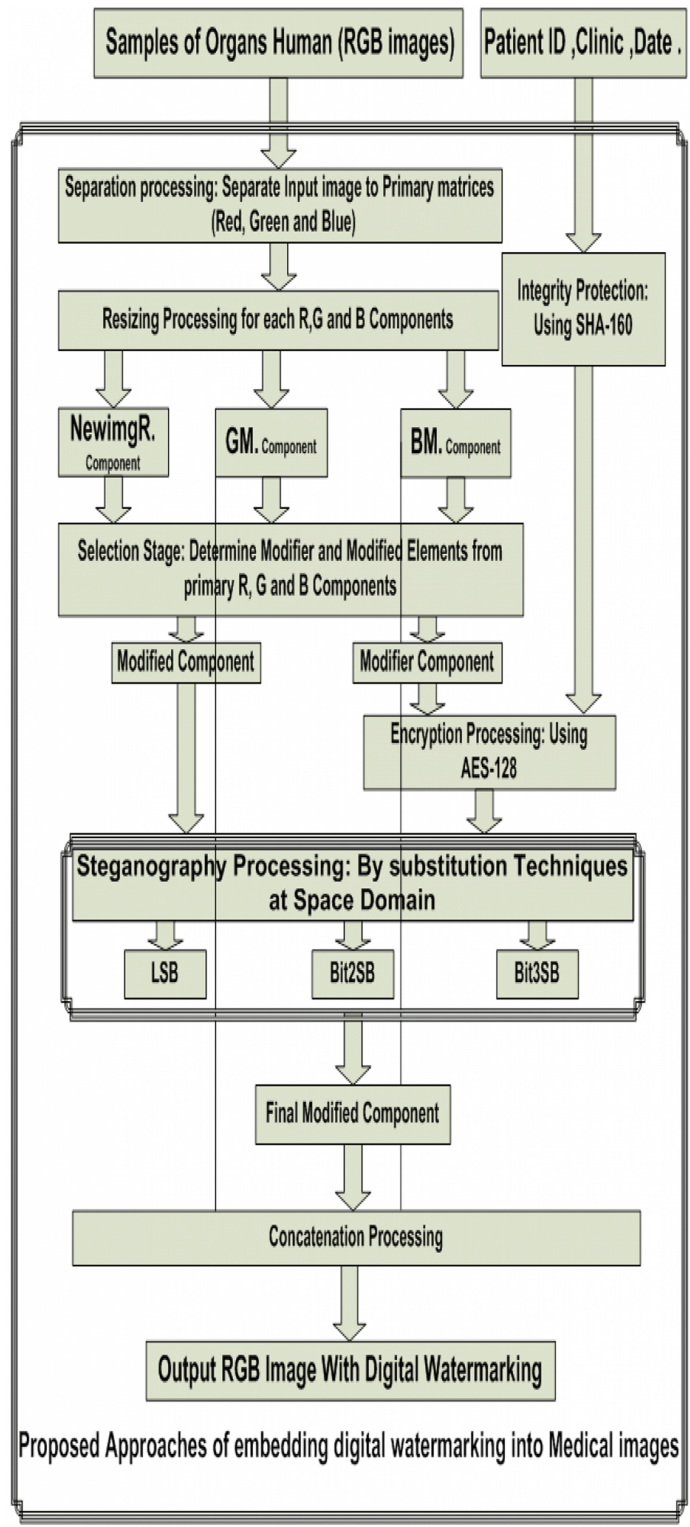

Figure 3 **Proposed approach of embedding digital watermarking into medical images.**
Full-size  DOI: 10.7717/peerj-cs.400/fig-3

The whole structure of the proposed algorithm is summarized below:

Step 1. Read the input RGB color image

$$I = (Irc) = \begin{bmatrix} I_{11} & \dots & I_1m \\ \vdots & \ddots & \vdots \\ In_1 & \dots & Inm \end{bmatrix} \tag{1}$$

Step 2. The Color Filter Array is used to decompose the image into three components as shown in (2).

$$G = (Grc) = \begin{bmatrix} G_{11} & \dots & G_1m \\ \vdots & \ddots & \vdots \\ Gn_1 & \dots & Gnm \end{bmatrix}$$

$$R = (Rrc) = \begin{bmatrix} R_{11} & \dots & R_1m \\ \vdots & \ddots & \vdots \\ Rn_1 & \dots & Rnm \end{bmatrix}$$

$$B = (Rrc) = \begin{bmatrix} B_{11} & \dots & B_1m \\ \vdots & \ddots & \vdots \\ Bn_1 & \dots & Bnm \end{bmatrix} \tag{2}$$

Step 3. The separated components are resized to be suitable for the encryption process.

$$\text{RM}(r) = R(r + (8 - (r\text{MOD}8))) \tag{3}$$

if remender of numeric value for input rows ≠ 0
else,

$$\text{RM}(r) = R(r)$$

$$\text{RM}(c) = R(c + (8 - (c\text{MOD}8))) \tag{4}$$

if remender of numeric value for input rows ≠ 0
else,

$$\text{RM}(c) = R(c)$$

Step 4: The blue component is selected to be the pre-modifier component.

Step 5. The red component is called the modified component and by using the dynamic keys approach, the blue component (the pre-modifier component) is encrypted. The arrival date, the name, the patient ID, and any information related to the patient are combined using the dynamic keys. SHA-1 is used to get the initial Key for the Symmetric Encryption process.

The initial hash values are:

H (0) ="67452301",

H [1] ="FCDAB89",

H [2] ="98BADCFE",

H [3] ="10325476",

H [4] ="C3D2E1F0",

$$K1 = SHA - 1(\text{patient data}) = 86E152C142DB1256FC1EF004ADEB7E5741D94D \quad (5)$$

The initial key for encryption processing is the first 16 digits from the output

The ciphered blue component of the original color image is presented by the CB and the Key K = "86E152C142DB1256" with 44 internal key and 80 round according to (6).

$$CBrc = AES - 128([K], (BMrc)) = \begin{bmatrix} CB^{11} & \dots & CB^{1}m \\ \vdots & \ddots & \vdots \\ CBn^{1} & \dots & CBnm \end{bmatrix} \quad (6)$$

Step 6. After the encryption is carried out, the new blue component is called the modifier component.

Step 7. To get the final modified component, a spatial domain substitution processing is carried out between the modified and modifier components.

$$NRM = (NRMrc \leftarrow CBrc) = \begin{bmatrix} NRM_{11} & \dots & NRM_{1m} \\ \vdots & \ddots & \vdots \\ NRMn_{1} & \dots & NRMnm \end{bmatrix} \leftarrow \begin{bmatrix} CB_{11} & \dots & CB_{1}m \\ \vdots & \ddots & \vdots \\ CB_{n1} & \dots & CBnm \end{bmatrix} \quad (7)$$

Step 8. The substitution Step is carried out between one of the Least Significant bits, bit number #2 and bit number #3 with the same bit of ciphered blue component according to (8, 9, and 10) to get the final modified component.

$$(NRMrc) = [b_7, b_6, b_5, b_4, b_3, b_2, b_1, b_0(CB)] \quad (8)$$

$$(NRMrc) = [b_7, b_6, b_5, b_4, b_3, b_2, b_1(CB), b_0] \quad (9)$$

$$(NRMrc) = [b_7, b_6, b_5, b_4, b_3, b_2(CB), b_1, b_0] \quad (10)$$

Step 9. The original blue and green components with the final modified component are used to compose the final secured image accordingly to (11), where NI is the protected image.

$$(NIrc) = \{(NRMrc), (GMrc), (BMrc)\} \quad (11)$$

# IMPLEMENTATION AND EXPERIMENTAL RESULTS

In the experiment, MATLAB 2015 Version 8.5.0.197613, core i3 processor 2.3 GHz and 4 GB RAM is used as the test platform.

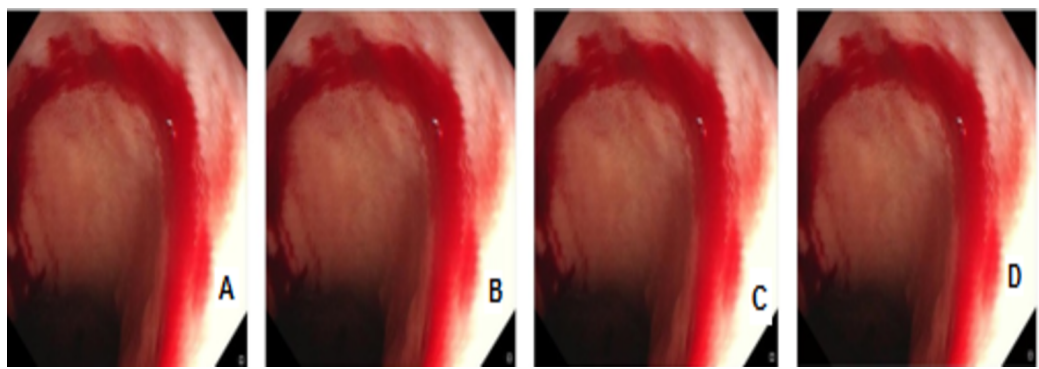

**Figure 4 Digital watermarking using different strategies for the AB image.** (A) AB image without digital watermarks (B) AB digital watermarked by LSB (C) AB digital watermarked by Bit2SB (D) AB digital watermarked by Bit3SB. 

In Figs. 4–9 randomly selected medical images are used as original images images For performance evaluation, the parameters have been calculated as follows:

## Embedding capacity

Although there is a watermark in every pixel in the image, it represents 1/8 of the pixel, but it did not affect the quality of the image.

## Image distortion

To assist the image distortion the MSE. MAE, the Structural Similarity Index (SSIM), and Universal Image Quality Index (UIQI) are used. The mean squared error (MSE) is calculating by averaging the squared intensity differences of the watermarked image and the reference image pixels. The (SSIM) is calculated by normalizing the mean value of structural similarity between the original image and the watermarked one. Based on the combination between the luminance distortion, the contrast distortion, and the loss of correlation, the UIQI is designed by modeling any distortion of the image.

$$\text{MSE} = \frac{1}{M*N} * \sum_{j=1}^{n} [x(i,j) - v(i,j)]^2 \tag{12}$$

$$\text{MAE} = \frac{1}{N} \sum_{i=1}^{N} |x_i - v_i| \tag{13}$$

$$\text{SSIM}(X, V) = (2 * \bar{X} * \bar{V} + c_1) * \frac{(2 * \sigma_{XV} + c_2)}{(\bar{x}^2 + \bar{v}^2 + c_1) * (\sigma_x^2 + \sigma_v^2 + c_2)} \tag{14}$$

$$Q = \left(\frac{\sigma_{xv}}{\sigma_x \sigma_v}\right) * \left(\frac{2 * \bar{x} * \bar{v}}{\bar{x}^2 + \bar{v}^2}\right) * \left(\frac{2 * \sigma_x * \sigma_v}{\sigma_x^2 + \sigma_v^2}\right) \tag{15}$$

Tables 1 and 2 shows that the mean absolute error and mean square error of the single LSB has the lowest value compared with the other elements.

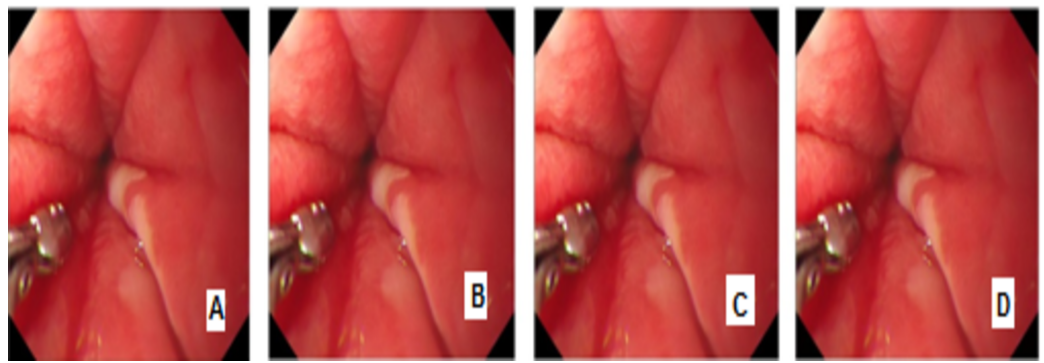

**Figure 5 Digital watermarking using different strategies for the AC image.** (A) AC image without digital watermarks (B) AC digital watermarked by LSB (C) AC digital watermarked by Bit2SB (D) AC digital watermarked by Bit3SB. Full-size ◨ DOI: 10.7717/peerj-cs.400/fig-5

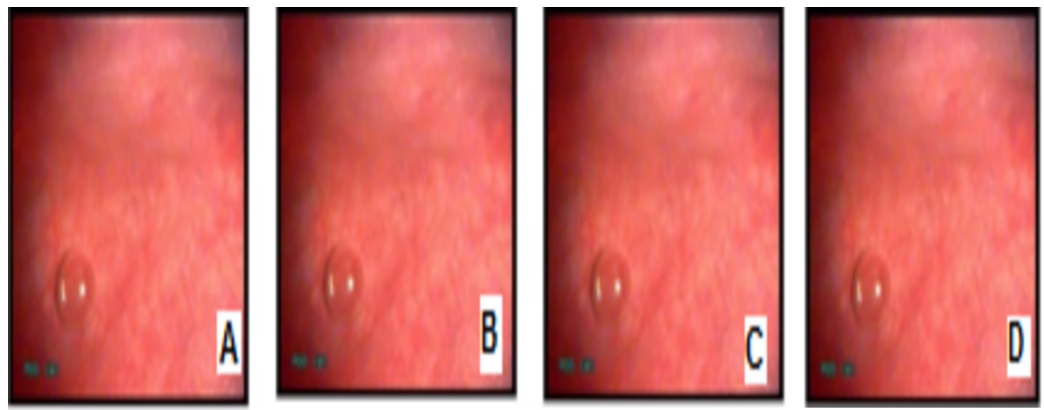

**Figure 6 Digital watermarking using different strategies for the Flu image.** (A) FLU image without digital watermarks (B) FLU digital watermarked by LSB (C) FLU digital watermarked by Bit2SB (D) FLU digital watermarked by Bit3SB. Full-size ◨ DOI: 10.7717/peerj-cs.400/fig-6

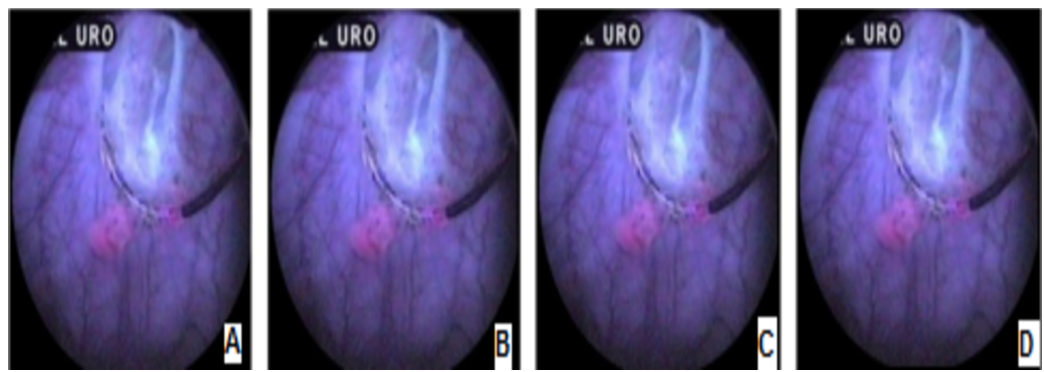

**Figure 7 Digital watermarking using different strategies for the AA image.** (A) AA image without digital watermarks (B) AA digital watermarked by LSB (C) AA digital watermarked by Bit2SB (D) AA digital watermarked by Bit3SB. Full-size ◨ DOI: 10.7717/peerj-cs.400/fig-7

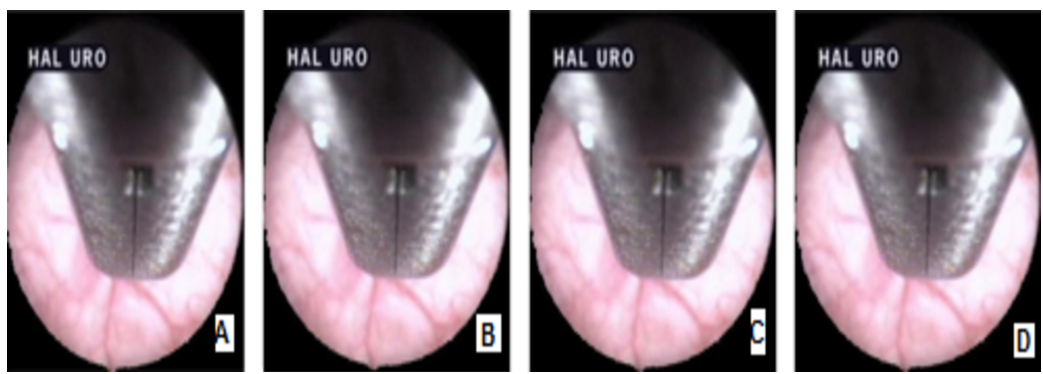

**Figure 8 Digital watermarking using different strategies for the Thr image.** (A) Thr image without digital watermarks (B) Thr digital watermarked by LSB (C) Thr digital watermarked by Bit2SB (D) Thr digital watermarked by Bit3SB. Full-size 🖼 DOI: 10.7717/peerj-cs.400/fig-8

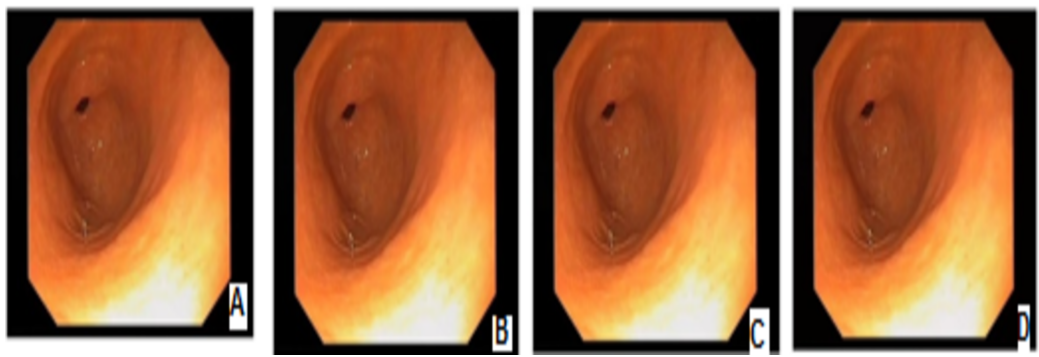

**Figure 9 Digital watermarking using different strategies for the Gastro image.** (A) Gastro image without digital watermarks (B) Gastro digital watermarked by LSB (C) Gastro digital watermarked by Bit2SB (D) Gastro digital watermarked by Bit3SB. Full-size 🖼 DOI: 10.7717/peerj-cs.400/fig-9

**Table 1 The mean absolute error of the proposed approach.**

| Images/size/JPEG | Mean absolute error of the proposed approach | | |
|---|---|---|---|
| | Single-LSB | Single-Bit2SB | Single-Bit3SB |
| Thr/472*499, 21.9 KB | 0.1025 | 0.182 | 0.4583 |
| AA/459*442, 21.9 KB | 0.1022 | 0.1837 | 0.4519 |
| AB/512*512, 22 KB | 0.0824 | 0.1611 | 0.3342 |
| AC/512*512, 340 KB | 0.0832 | 0.1643 | 0.3369 |
| Flu/348*288, 26 KB | 0.0827 | 0.1608 | 0.3205 |
| Gastro/640*480, 47.1 KB | 0.0968 | 0.1654 | 0.4456 |

Table 3 shows the Structural Similarity Index of the single LSB and single-Bit2SB have the highest value compared with the single-Bit3SB.

Table 4 shows that the proposed approach has a high Entropy value for all elements.

**Table 2** The mean square error of the proposed approach.

| Images/size/JPEG | Mean square error of the proposed approach | | |
|---|---|---|---|
| | Single-LSB | Single-Bit2SB | Single-Bit3SB |
| Thr/472*499, 21.9 KB | 0.171390178 | 0.636176128 | 2.905030939 |
| AA/459*442, 21.9 KB | 0.170218532 | 0.63784765 | 2.877668309 |
| AB/512*512, 22 KB | 0.167069753 | 0.664449056 | 2.700602214 |
| AC/512*512, 340 KB | 0.166893005 | 0.663869222 | 2.703511556 |
| Flu/348*288, 26 KB | 0.167730638 | 0.665400752 | 2.672935957 |
| Gastro/640*480, 47.1 KB | 0.168408203 | 0.649548611 | 2.869375 |
| Thr/472*499, 21.9 KB | 0.171390178 | 0.636176128 | 2.905030939 |

**Table 3** The SSIM for the proposed approach.

| Images/size/JPEG | SSIM of the proposed approach | | |
|---|---|---|---|
| | Single-LSB | Single-Bit2SB | Single-Bit3SB |
| Thr/472*499, 21.9 KB | 0.996973577 | 0.991946693 | 0.945857695 |
| AA/459*442, 21.9 KB | 0.997807567 | 0.994891188 | 0.959116353 |
| AB/512*512, 22 KB | 0.999295797 | 0.997930044 | 0.987737559 |
| AC/512*512, 340 KB | 0.999673577 | 0.999171438 | 0.993768634 |
| Flu/348*288, 26 KB | 0.999867319 | 0.999476348 | 0.997853068 |
| Gastro/640*480, 47.1 KB | 0.997975423 | 0.993904738 | 0.961204255 |

**Table 4** The UIQI for the proposed approach.

| Images/size/JPEG | UIQI of the proposed approach | | |
|---|---|---|---|
| | Single-LSB | Single-Bit2SB | Single-Bit3SB |
| Thr/472*499, 21.9 KB | 0.999974809 | 0.999903907 | 0.999592958 |
| AA/459*442, 21.9 KB | 0.999910199 | 0.996653494 | 0.998524767 |
| AB/512*512, 22 KB | 0.999952895 | 0.999812608 | 0.999240437 |
| AC/512*512, 340 KB | 0.999891903 | 0.99956814 | 0.998248895 |
| Flu/348*288, 26 KB | 0.999951511 | 0.999807683 | 0.999228211 |
| Gastro/640*480, 47.1 KB | 0.999973065 | 0.999891465 | 0.999576599 |

## Security of the watermark

The patient ID and any information related to the patient are combined using the dynamic keys. SHA-1 is used to get the initial Key for the Symmetric Encryption process. The key is dynamic and depends on the patient information, this makes it easier to detect intentional changes.

**Table 5  The peak signal to noise ratio of the proposed approach.**

| Images/size/JPEG | Peak signal to noise ratio of the proposed approach | | |
|---|---|---|---|
| | Single-LSB | Single-Bit2SB | Single-Bit3SB |
| Thr/472*499, 21.9 KB | 55.79094431 | 50.09502993 | 43.49929599 |
| AA/459*442, 21.9 KB | 55.8207352 | 50.08363401 | 43.54039627 |
| AB/512*512, 22 KB | 55.9018253 | 49.90618672 | 43.81619742 |
| AC/512*512, 340 KB | 55.90642225 | 49.90997826 | 43.81152131 |
| Flu/348*288, 26 KB | 55.88467963 | 49.89997073 | 43.86091808 |
| Gastro/640*480, 47.1 KB | 55.86717119 | 50.00468702 | 43.55293051 |

**Table 6  The peak signal to noise ratio of some of the existing approaches.**

| Techniques | PSNR (dB) |
|---|---|
| *Alattar (2004)* | 22.36 |
| *Shih & Wu (2005)* | 29.23 |
| *Thodi & Rodríguez (2007)* | 38.0 |
| *Zhao et al. (2011)* | 29.39 |
| *Wang, Lee & Chang (2013)* | 44.64 |
| *Shih & Zhong (2016)* | 51.24 |
| *Bamal & Kasana (2018)* | 48.53 |

## Invisibility evaluation

The PSNR between the watermarked image and the original image is used to evaluate the visual quality.

$$PSNR = 10\log_{10}\left(\frac{S^2}{MSE}\right) \qquad (16)$$

Table 5 shows that the Single-LSB has a better result compared with the Single-Bit2SB and the Single-Bit3SB. Table 6 shows the peak signal to noise ratio of some of the existing approaches. The results show that the PSNR of the proposed approach is better than the tested approaches.

## COMPARISON RESULTS

By comparing the proposed work with the existing work, the created code is distributed at every pixel of the cover medical image and not only in the regions of non-interest pixels, while the image details are still preserved. Most of the existing watermark is embedded in non-interest, which facilitates the attacking process. The proposed algorithm succeeded in determining the intentional changes in the patient image up to 2*2 pixels. There is a big difference between the proposed model and the existing models which is the performance of the existing approach measured by its ability to extract the watermark from the original image. But the proposed approach preserves the watermark and the original image at the same time. This difference gives the proposed model the advantage of discovering the attacks that the images may be exposed to. This is because the proposed method creates the watermark from one of the components of the original image and not by adding a watermark image to the image.

## DISCUSSIONS

This subsection discusses the proposed algorithm in various aspects such as Authenticity and integrity, and prevention against intentional attacks.

### Authenticity and integrity

Authenticity and security are the most significant requirements for medical image protection. Because with any deliberate change of the image, a wrong decision by the physician will result.

Figure 10 demonstrates the proposed algorithm that detects the attacks:

The vulnerability assessment approach is based on assuming intentional changes in the patient image up to 2*2 pixels.

The evaluation also includes the change between the color components. Figures (11)–(14) demonstrate different attack regions, and all of them are detected by the proposed algorithm.

Table 7 shows the ability of the proposed algorithm to detect the intentional attacks.

From the image quality point of view, it is clear that the choice of the single LSB gave the best image quality performance over the rest of the elements and the single-Bit2SB comes in the second rank.

From the ability of the algorithm to detect the intentional attacks, the proposed algorithm succeeded in determining intentional changes in the patient image up to 2*2 pixels especially in the case of Single-Bit2SB.

## FUTURE WORK

The aim of embedding the watermark is to preserve the medical images from any intentional or unintended interference. Therefore, in the future, we will try to find different methods to achieve this goal and for that, we propose the following:

1. To protect the blockchain record, a new idea will be proposed based on the combination of the watermarking, the QR code, and the Interplanetary File System (IPFS).

2. Use of word embedding (*Smith, 2019*) technique to create an efficient watermark. Word embedding is one of the important developments in natural language processing. The key benefit of word embedding is that by preserving the semantic similarity of words and constructing a low-dimensional vector, it allows for a more expressive and effective representation.

3. The t-SNE dimensionality reduction (*Heuer, 2016*) can be used to compare the original image with the watermarked image and it will help in the extraction of the watermark.

4. The amount of cybersecurity data generated in the form of unstructured texts (*Simran et al., 2019*), such as social media tools, blogs, posts, and so on, has increased remarkably in recent years. Called Entity Recognition (NER) is an initial step in the conversion of this unstructured data into structured data that many applications may use. The unstructured texts could be used to create a watermark that would be difficult to expose to deliberate attacks that would alter patients' information

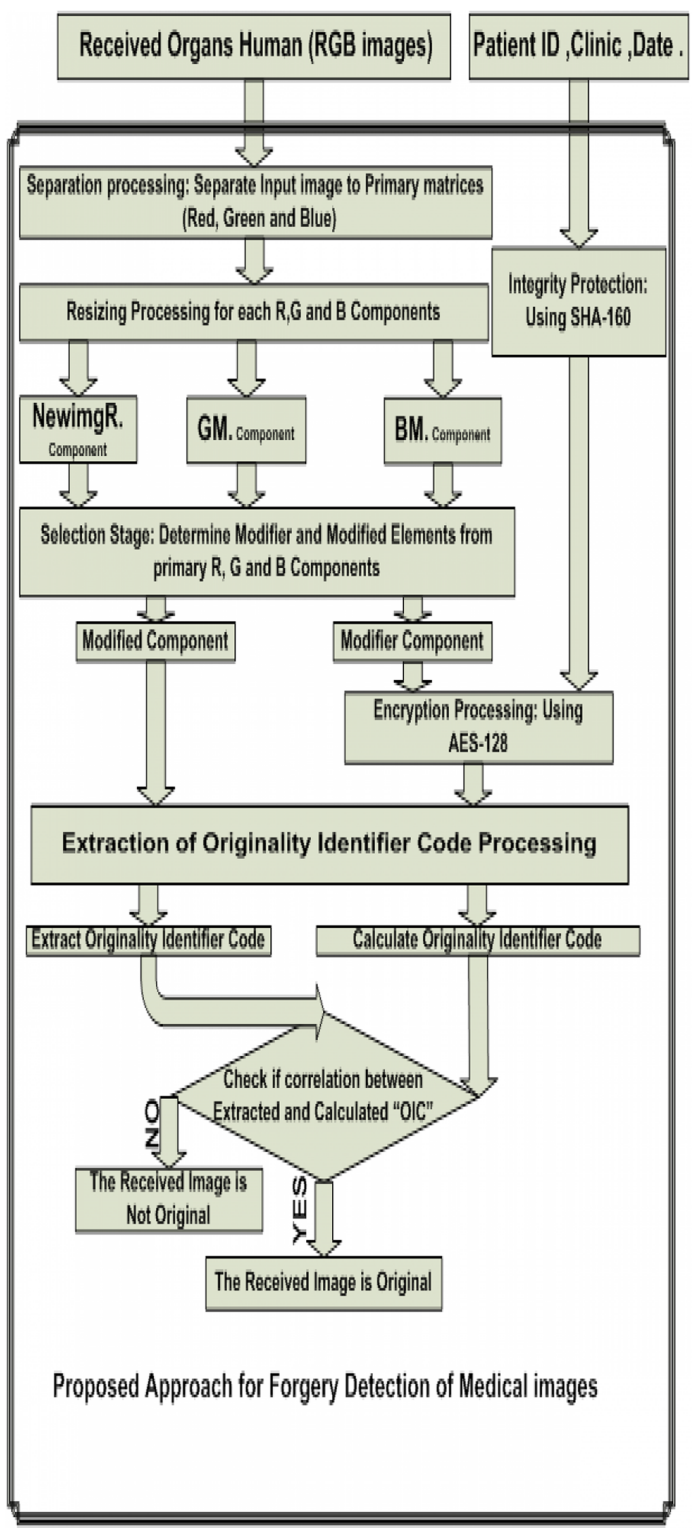

Proposed Approach for Forgery Detection of Medical images

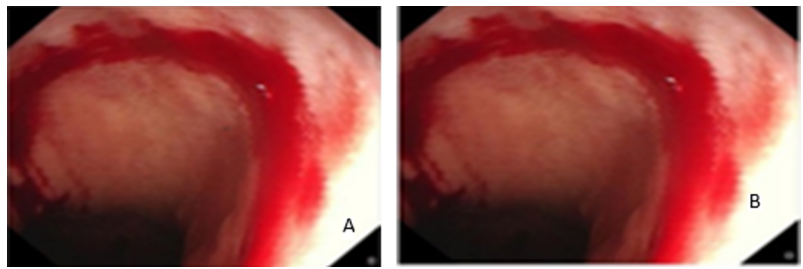

**Figure 11 The modification of the AB image.** (A) The AB image is modified at red the by same blue component from raw 238–241 and column 300–303 (B) the AB image is modified in Red by Blue component at row from 383 to 386 and column 313–316 LSB.

Full-size 🖼 DOI: 10.7717/peerj-cs.400/fig-11

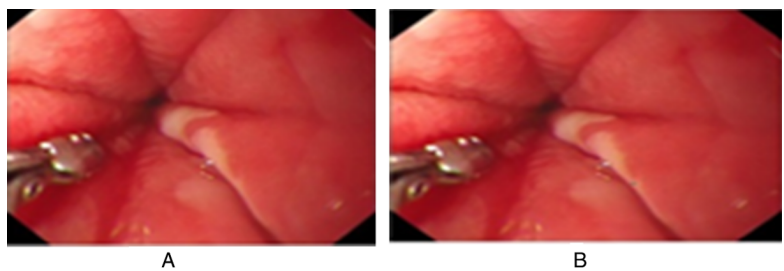

**Figure 12 The modification of the AC image.** (A) The AC image is modified at blue the by same green component from raw 238–241 and column 300–303 (B) the AC image is modified in Red component by 1 at row from 383 to 386 and column 313–316 LSB. Full-size 🖼 DOI: 10.7717/peerj-cs.400/fig-12

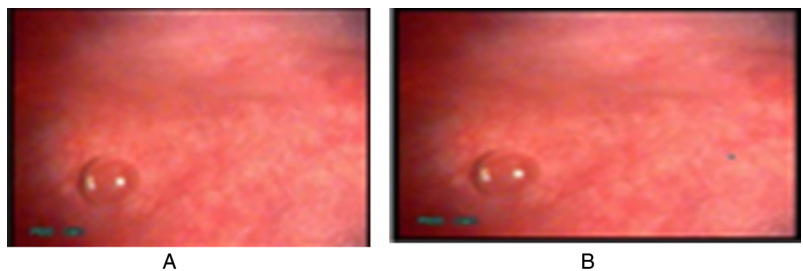

**Figure 13 The modification of the FLU image.** (A) The FLU image is modified at blue by green component at row from 138 to 141 and column 200–203 (B) the FLU image is modified at red by the same blue component at row from 183 to 186 and column 316–319.

Full-size 🖼 DOI: 10.7717/peerj-cs.400/fig-13

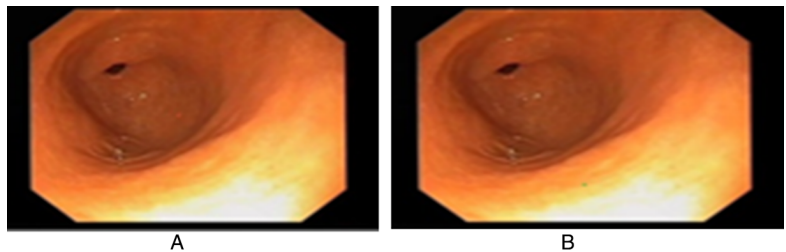

**Figure 14 The modification of the Gastro image.** (A) The Gastro image is modified at red component by 255 in raw 238–241 and column 300–303 (B) the Gastro image is modified in Red by Blue component at row from 383 to 386 and column 313–316 LSB. Full-size 🖼 DOI: 10.7717/peerj-cs.400/fig-14

**Table 7** The forgery detection rate.

| Tampared block size to image (TBSI) % | X "Forgery Detection Rate (FDR) based on correlation between original image and Tampared image" | |
|---|---|---|
| Thr, (4*4)/(472*499) = 0.00679 | Successful | 0.999778433 |
| AA, (4*4)/(459*442) = 0.0078865 | Successful | 0.999927097 |
| AB, (4*4)/(512*512) = 0.006104 | Successful | 0.9999840461 |
| AC(4*4)/(512*512) = 0.006104 | Successful | 0.999955139 |
| Flu, (4*4)/(348*288) = 0.015964 | Successful | 0.999123523 |
| Gastro, (4*4)/(640*480) = 0.005208 | Successful | 0.999819384 |

# CONCLUSIONS

In this article, a robust spatial watermarking algorithm for medical images has been proposed. This new approach to maintain high data security combines integrity protection, encryption algorithms, and steganography techniques.

The performance of the proposed algorithm is evaluated for the desired outcome is obtained without significant degradation in the extracted covered image and the ability of the code against the intentional destruction of the medical image.

The proposed approach has the following advantages:

- The processed images have the same size as input images.
- The proposed approach is valid for all the image shapes and not only for the rectangular shape as in many of the existing works.
- The approach applies to every RGB Images.
- The created secret code has dynamic properties based on:-Imaging device manufacturers. -basic components of color images (R, G, and B).
- Its advantage over the other approaches is that the created code is distributed at every pixel of the cover medical image and not only in the regions of non-interest pixels (RONI).
- Although the digital watermark is distributed on every pixel on the image the code does not affect the output image quality.
- The encryption of watermarks uses hash function properties, which are extremely sensitive to the initial value, to enhance the security of watermarks.
- Although there is a watermark in every pixel in the image, it represents 1/8 of the pixel, but it did not affect the quality of the image.
- It solves the problem that exists in the spatial domain algorithms that cannot simultaneously offer protection against intentional attacks and robustness.
- The proposed approach preserves the watermark and the original image at the same time. This is because the proposed method creates the watermark from one of the components of the original image and not by embedding a watermark image in the image.

## ADDITIONAL INFORMATION AND DECLARATIONS

### Funding
The authors received no funding for this work.

### Competing Interests
The authors declare that they have no competing interests.

### Author Contributions
- Ahmad Nagm conceived and designed the experiments, performed the computation work, authored or reviewed drafts of the paper, and approved the final draft.
- Mohammed Safy Elwan performed the experiments, analyzed the data, prepared figures and/or tables, and approved the final draft.

### Data Availability
The following information was supplied regarding data availability:

MATLAB code is available as a Supplemental File.

Publicly available data located at Mendeley:

Ali, Sharib; Zhou, Felix; Daul, Christian; Braden, Barbara; Bailey, Adam; East, James; Realdon, Stefano; Georges, Wagnieres; Loshchenov, Maxim; Blondel, Walter; Grisan, Enrico; Rittscher, Jens (2019), "Endoscopy Artefact Detection (EAD) Dataset", Mendeley Data, V1, DOI 10.17632/c7fjbxcgj9.1.

### Supplemental Information
Supplemental information for this article can be found online at http://dx.doi.org/10.7717/peerj-cs.400#supplemental-information.